\documentclass[12pt,preprint!]{aastex}

\usepackage{psfig}

\newcommand\pp     {$\pm$}
\newcommand\pers     {s$^{-1}$}

\newcommand\Lunit {erg s$^{-1}$}
\newcommand\funit {erg s$^{-1}$ cm$^{-2}$}

\def\degr{\hbox{$^\circ$}}

\righthead{BeppoSAX observations of SAX J1808.4--3658}
\slugcomment{Accepted for publication by ApJ, 30 January
2002}

\begin{document}

\title{Very low luminosities from the accretion-driven millisecond
X-ray pulsar SAX J1808.4--3658 during quiescence}

\author{Rudy Wijnands\altaffilmark{1,6}, Lucien
Kuiper\altaffilmark{2}, Jean in 't Zand\altaffilmark{3,2}, Tadayasu
Dotani\altaffilmark{4}, Michiel van der Klis\altaffilmark{5}, John
Heise\altaffilmark{2}}

\altaffiltext{1}{Center for Space Research, Massachusetts Institute of
Technology, 77 Massachusetts Avenue, Cambridge, MA 02139-4307, USA;
rudy@space.mit.edu}

\altaffiltext{2}{SRON Laboratory for Space Research, Sorbonnelaan 2,
NL-3584 CA, Utrecht, The Netherlands}

\altaffiltext{3}{Astronomical Institute, Utrecht University, P.O. Box
80000, NL-3507 TA, Utrecht, The Netherlands}

\altaffiltext{4}{Institute of Space and Astronautical Science, 3-1-1,
Yoshinodai, Sagamihara, Kanagawa 229-8510, Japan}

\altaffiltext{5}{Astronomical Institute ``Anton Pannekoek'',
       University of Amsterdam, Kruislaan 403, NL-1098 SJ Amsterdam,
       The Netherlands}

\altaffiltext{6}{Chandra Fellow}

\begin{abstract}

We have observed the millisecond X-ray pulsar SAX J1808.4--3658 on
three occasions during its 2000 outburst with the {\it BeppoSAX}
satellite. The source was highly variable and erratic during this
outburst, and by coincidence we obtained data only during times when
the source had very low luminosities.  During our observations, we
detected four faint sources. The source closest to the position of SAX
J1808.4--3658 is still $\sim$1.6$'$ away. This source can only be
identified with SAX J1808.4--3658 if we assume that the {\it BeppoSAX}
positional reconstruction is not completely understood.  We also
reanalyzed a {\it BeppoSAX} observation taken in March 1999 when the
source was in quiescence and during which the source was thought to
have been detected (Stella et al. 2000).  Based on the similarities
(position and luminosity) of this source with the above mentioned
source $\sim1.6'$ away from SAX J1808.4--3658, it is possible that
they are the same source. If this source is not the millisecond
pulsar, then during all {\it BeppoSAX} observations of SAX
J1808.4--3658 (the 2000 outburst ones and the 1999 quiescent one), the
millisecond pulsar was not detected. A reanalysis of the {\it ASCA}
quiescent data of SAX J1808.4--3658 (Dotani, Asai, \& Wijnands 2000)
confirms that during this observation the source was securely detected
in quiescence.  We discuss our results for SAX J1808.4--3658 in the
context of the quiescent properties of low-mass X-ray binary
transients.

\end{abstract}


\section{Introduction\label{section:introduction}}

In 1996 September, a new X-ray transient was discovered with the Wide
Field Cameras (WFCs) on board the {\it BeppoSAX} satellite and the
source was designated SAX J1808.4--3658 (In 't Zand et al. 1998). Type
I X-ray bursts were detected demonstrating that the compact object in
this system is a neutron star and that the source is a low-mass X-ray
binary (LMXB). From the bursts, an initial distance estimate of 4 kpc
was determined but this was later revised to 2.5 kpc (In 't Zand et
al. 1998, 2001). The maximum luminosity of SAX J1808.4--3658 was only
approximately $10^{36}$ \Lunit~ indicating that the source was a faint
neutron-star X-ray transient.  The outburst lasted about three weeks
after which the source returned to quiescence.  In April 1998, the
{\it Rossi X-ray Timing Explorer} ({\it RXTE}) satellite scanned,
using the Proportional Count Array (PCA), the region on the sky where
SAX J1808.4--3658 is located and the source was found to be active
again (Marshall 1998). Follow up {\it RXTE}/PCA observations showed
that SAX J1808.4--3658 exhibits coherent millisecond X-ray
oscillations with a frequency of approximately 401 Hz (Wijnands \& van
der Klis 1998), making this source the first (and so far the only)
accretion-powered millisecond X-ray pulsar discovered. The source was
also detected in the optical (Giles, Hill, \& Greenhill 1999, with an
optical position of R.A. = 18$^{\rm h}$ 08$^{\rm m}$ 27.54$^{\rm
s}$\pp0.015$^{\rm s}$ and Dec. --36\degr~58$'$ 44.3$''$\pp0.2$''$; all
coordinates in the paper are for J2000.0) and in the radio (Gaensler,
Stappers, \& Getts 1999).  Also this outburst lasted only for a few
weeks (e.g., Gilfanov et al. 1998).

During quiescence, SAX J1808.4--3658 was observed with the Narrow
Field Instruments (NFI) on board {\it BeppoSAX} and near the position
of SAX J1808.4--3658 a faint (a few times $10^{32}$ \Lunit) X-ray
source was discovered (Stella et al. 2000), which was identified with
the pulsar based on presumed positional coincidence. An {\it ASCA}
observation about half a year later resulted in a detection for SAX
J1808.4--3658 in quiescence but with a luminosity at least 4 times
lower than measured with {\it BeppoSAX} (Dotani, Asai, Wijnands 2000).
This low luminosity makes SAX J1808.4--3658 the dimmest quiescent
neutron-star LMXB X-ray transient detected so far (see, e.g., Asai et
al. 1996, 1998).

SAX J1808.4--3658 stayed dormant until 2000 January 23, when {\it
RXTE} found the source to be active again (van der Klis et al
2000). Due to solar constraints, the source was unobservable before
this date (from the end of November 1999) and it is unclear when
exactly the source had become active.  The source was only detected at
low luminosities (up to a few times $10^{35}$ \Lunit) with highly
variable activity lasting up to 13 May 2000 (Wijnands et al. 2000,
2001).  Here we report on {\it BeppoSAX}/NFI observations during this
outburst.

\section{The 2000 March {\itshape BeppoSAX}/NFI observations 
\label{section:2000march}}

To study the 0.2--100 keV X-ray spectrum of SAX J1808.4--3658, we had
{\it BeppoSAX}/NFI target-of-opportunity observations planned during
its anticipated 2000 outburst. These observations were triggered and
we obtained 3 observations in March 2000 (see Tab.~\ref{tab:log}; for
a total of 89.1 ksec). The data were analyzed using several different
approaches, but here we report only on the maximum-likelihood ratio
(MLR) method which is the most sensitive one (the results of the other
approaches were consistent with the results obtained from the MLR
analysis). The details of this method are described in Kuiper et
al. (1998) and In 't Zand et al. (2000; see also, e.g., Kuulkers et
al. 2000). As already mentioned above in
section~\ref{section:introduction}, the source was highly variable
during the 2000 outburst (see Wijnands et al. 2001) and unfortunately
all our {\it BeppoSAX}/NFI observations were taken during times when
the source had very low luminosities and consequently only the data
obtained with the MECS were useful.

\begin{figure}
\begin{center}
\begin{tabular}{c}
\psfig{figure=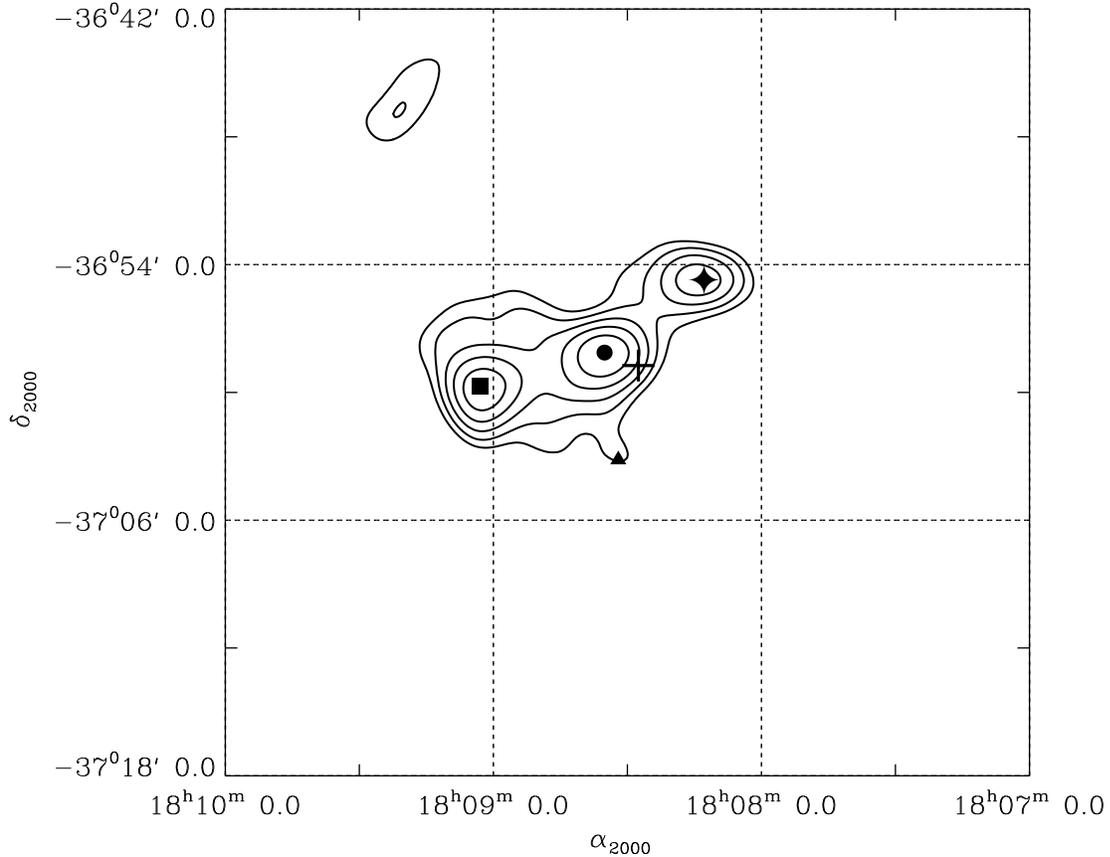,width=14cm}
\end{tabular}
\figcaption{\label{fig:image} Maximum likelihood ratio image of the
three 2000 March {\it BeppoSAX}/NFI/MECS observations combined (for
1.3--10 keV). The contour levels represent the source detection
significance atop an isotropic (flat) background starting at a
5$\sigma$ significance level in steps of 1$\sigma$ (assuming 1 degree
of freedom; see Kuiper et al. 1998 and In 't Zand et al. 2000 for the
procedure used to generate these images).  Four sources are clearly
detected ($>5\sigma$) in the center of the image. The position of the
millisecond X-ray pulsar is indicated with a plus. The position of the
four detected {\it BeppoSAX} sources are indicated by the filled
symbols (the {\it BeppoSAX} source: circle; SAX J1809.0--3659: square;
SAX J1808.2--3654: diamond; SAX J1808.5--3703: triangle) }
\end{center}
\end{figure}

To optimize the sensitivity towards detecting sources, we combined the
data of the three observations.  The resulting MLR image is shown in
Figure~\ref{fig:image}. In the center of this image four sources are
detected; their coordinates are listed in Table~\ref{tab:fov} (using
the latest spatial calibration; Perri \& Capalbi 2001, private
communication; the positions obtained by using the three observations
individually varied only by $\sim$10$''$).  The brightest source in
the field is the one closest to the position of SAX
J1808.4--3658. Sufficient photons were detected from this source to
extract its spectrum, which could satisfactorily be modeled with a
power-law with photon index of 1.65\pp0.10. Assuming a column density
$N_{\rm H}$ of 1.22 $\times 10^{21}$ cm$^{-2}$ (using the $A_{\rm v}$
from Wang et al. 2001 and the relation between $N_{\rm H}$ and $A_{\rm
v}$ from Predehl \& Schmitt 1995), the unabsorbed fluxes are 1.2\pp0.2
$\times10^{-13}$ (0.5--5 keV) and 1.9\pp0.3 $\times10^{-13}$ (0.5--10
keV) \funit.  Using a black-body, we obtained a temperature $kT$ of
1.05\pp0.03 keV and unabsorbed fluxes of 0.9\pp0.2 $\times10^{-13}$
(0.5--5 keV) and 1.2\pp0.2 $\times10^{-13}$ (0.5--10 keV)
\funit. However, the power-law fit is preferred over the black-body
fit (reduced $\chi^{2}$ is 1.1 vs. 1.9, respectively). When fitting a
two component model (e.g., power-law plus a black-body), the fit
parameters could not usefully be constrained.

The position of this brightest source, however, is located 1.63$'$
from the position of the optical counterpart of the millisecond X-ray
pulsar, which is considerably larger than the 0.55$'$ error radius
(68\% confidence level (1$\sigma$); conservatively calculated as the
direct sum of the statistical (18$''$) and systematical (15$''$)
1$\sigma$ errors) on the source position.  The millisecond X-ray
pulsar is located at the 3$\sigma$ location contour of this brightest
{\it BeppoSAX} source, and the probability that this source can still
be identified with the millisecond X-ray pulsar is
$\sim$0.0029. Although this would indicate that it is unlikely that
this source can be identified with the millisecond pulsar, it is
possible that the systematics in the position reconstruction for {\it
BeppoSAX} are not fully understood and we cannot rule out completely
the possibility that this source is indeed SAX J1808.4--3658. Until it
is conclusively demonstrated that this source is the pulsar or not, we
will refer to this source as ``the {\it BeppoSAX} source''.

\section{The 1999 March {\itshape BeppoSAX}/NFI observation}
 
We have reanalyzed the 1999 March 17--19 {\it BeppoSAX}/NFI quiescent
observation of SAX J1808.4--3658.  We confirm the presence of the
source reported by Stella et al. (2000), however, the position of this
source (R.A. = 18$^{\rm h}$ 08$^{\rm m}$ 32.9$^{\rm s}$; Dec. =
--36\degr~58$'$ 07.5$''$; error radius 0.95$'$, 68\% confidence level)
and its count rates (2.0\pp0.5 $\times 10^{-3}$ counts \pers;
1.3--10.0 keV; see also Stella et al. 2000 whom obtained very similar
coordinates and count rates) are almost identical to those obtained
for the {\it BeppoSAX} source during the combined 2000 March
observations, suggesting that it is the same source (the detected
source is $\sim$1.2$'$ away from the position of the pulsar).  If this
source cannot be identified with the pulsar, then during all the {\it
BeppoSAX}/NFI observations (both the 1999 quiescent observation and
the 2000 outburst ones), the millisecond X-ray pulsar was not
detected.

\section{The 1999 September \itshape{ASCA} observation \label{section:asca}}

\begin{figure}
\begin{center}
\begin{tabular}{c}
\psfig{figure=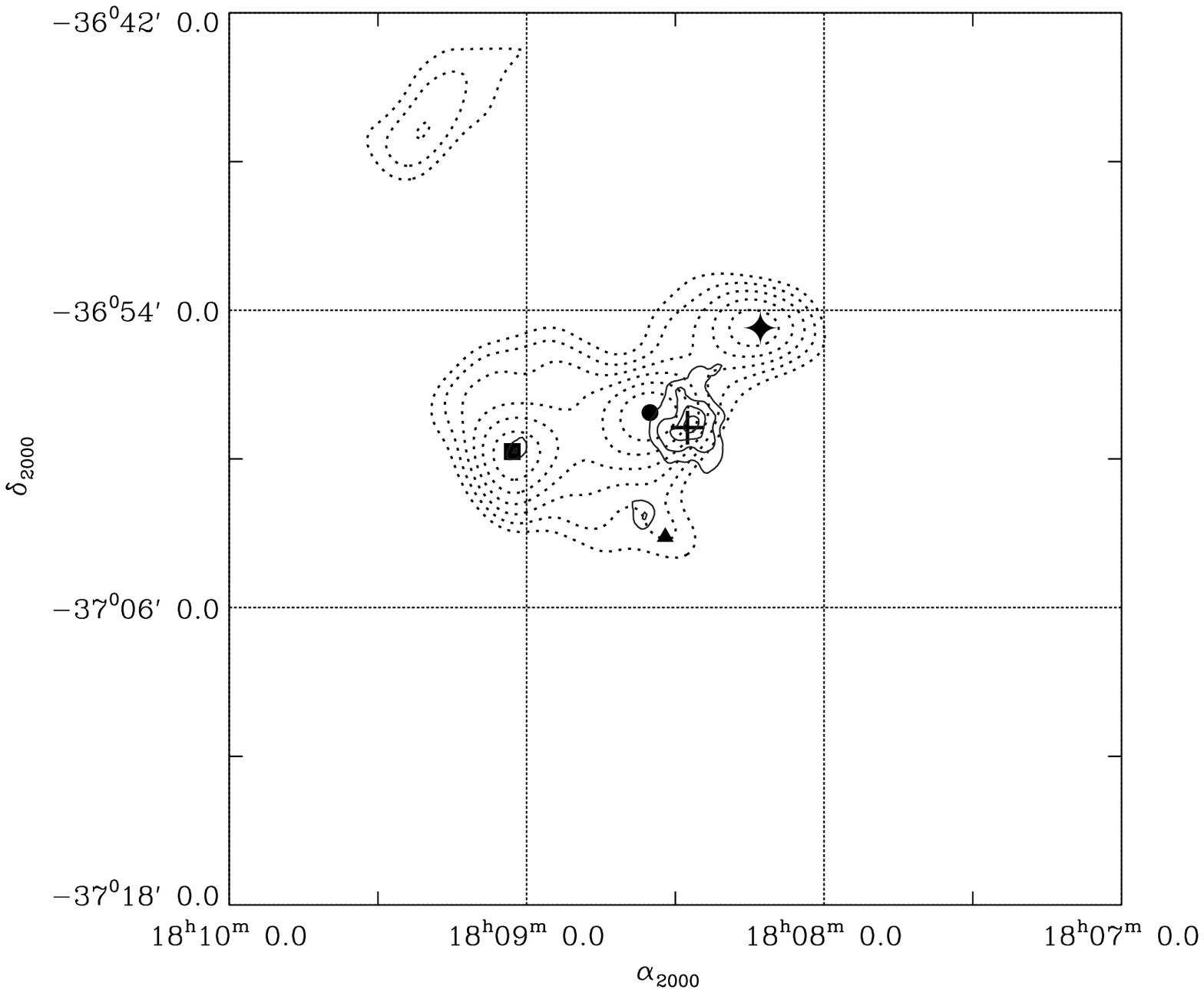,width=8cm}~~~~~~~\psfig{figure=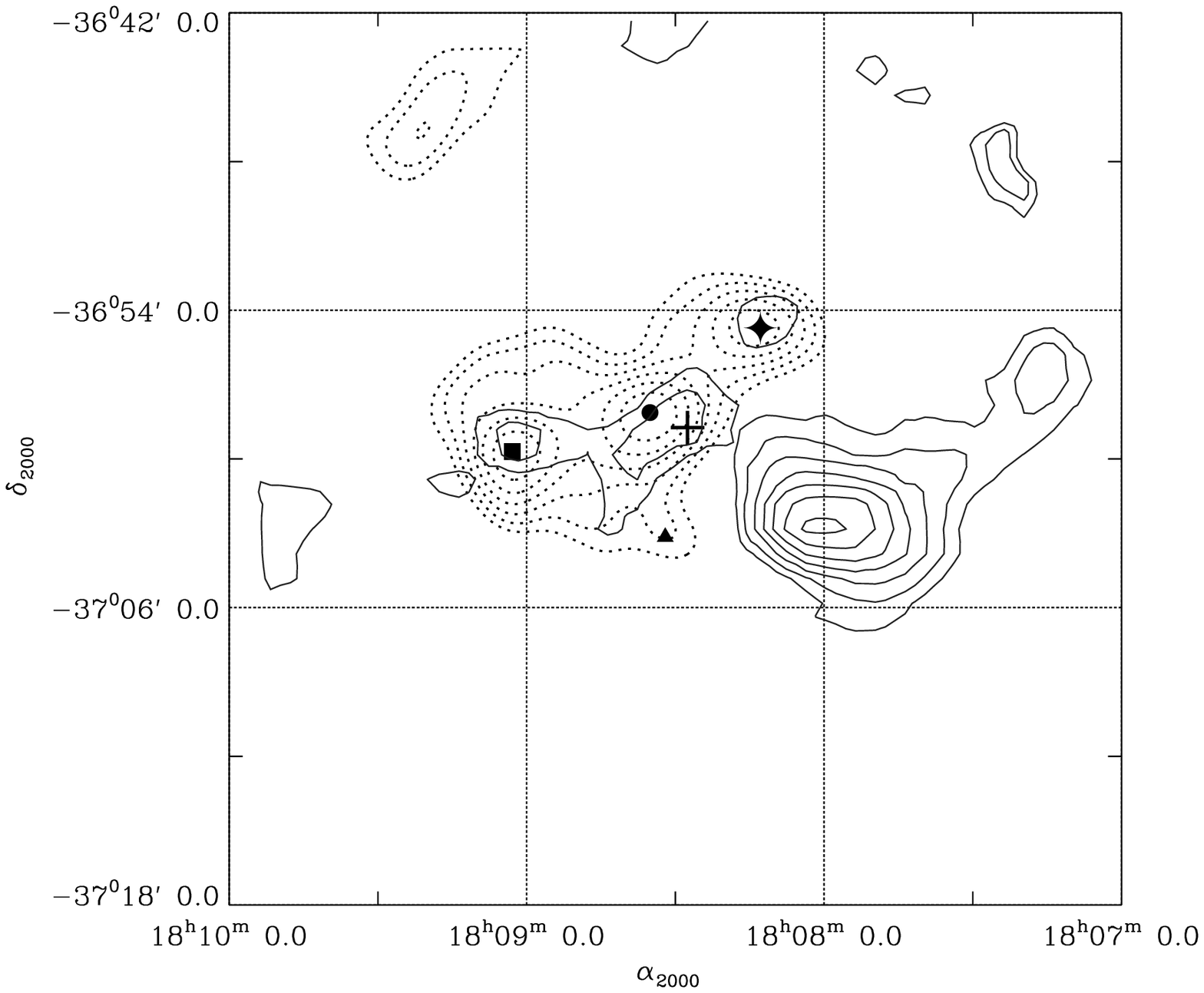,width=8cm}
\end{tabular}
\figcaption{\label{fig:ascasis_image} The maximum likelihood ratio
images of the {\it ASCA} data (solid lines; SIS: {\it left} panel;
1.3--10 keV; GIS: {\it right} panel; no energy selection) over-plotted
on the 1.3--10 keV {\it BeppoSAX} data (dashed lines; see also
Fig.~\ref{fig:image}).  The contour levels represent the source
detection significance atop an isotropic (flat) background starting at
a 4$\sigma$ level (for {\it BeppoSAX}/MECS and {\it ASCA}/SIS;
3$\sigma$ for {\it ASCA}/GIS) in steps of 1$\sigma$ assuming 1 degree
of freedom. For the symbols see Figure~\ref{fig:image}.}
\end{center}
\end{figure}

We have reanalyzed the quiescent {\it ASCA} data of SAX J1808.4--3658
(Dotani et al. 2000; see Tab.~\ref{tab:log}) to investigate if the
four detected {\it BeppoSAX} sources are also detected with {\it
ASCA}. We used two different methods to analyze the {\it ASCA} data;
the method applied by Dotani et al. (2000; hereafter referred to as
the standard method), but also the MLR method. The morphology of the
{\it ASCA}/SIS images obtained with both methods (see
Fig.~\ref{fig:ascasis_image} for the MLR image and Dotani et al. 2000
for the image obtained with the standard method) are very similar to
each other, and a source was detected in the center of the images (at
$\sim8\sigma$ significance, using the MLR method) and its position is
consistent with the position of SAX J1808.4--3658 (see
Tab.~\ref{tab:fov}; using an updated spatial calibration of {\it
ASCA}; Gotthelf et al. 2000; we only list the positions obtained with
the MLR method).  We also applied the MLR method to the combined GIS2
and GIS3 data. Although the coarse image binning and relatively high
background prevented us from applying the standard analysis (Dotani et
al. 2000), the MLR method may be better suited to reconstruct such a
noisy, coarse-rebinned image. The resulting MLR image is shown in
Figure~\ref{fig:ascasis_image}.  We could detect the millisecond X-ray
pulsar at a significance of $\sim 5\sigma$.

The {\it ASCA}/SIS count rate (0.5--5.0 keV) obtained using the MLR
method is 2.4\pp0.3 $\times 10^{-3}$ counts \pers, which is
significantly higher than the one we obtained using the standard
method (1.1\pp0.4 $\times 10^{-3}$ counts \pers). However, for faint
sources it is important to correctly estimate the local background
level at the source position, which is not trivial due to the extended
{\it ASCA} point spread-function.  To investigate the effect of a more
structured background on the source count rate, we modeled the
background near the pulsar in terms of including 3 extra point sources
at ``hot'' positions in the MLR image, which are defined as the
positions of remaining excesses in a residual MLR image after the
detected sources were modeled out. Using this background model, the
count rate of the millisecond X-ray pulsar decreases to 1.6\pp0.3
$\times 10^{-3}$ counts \pers. This count rate is still $\sim$50\%
higher than obtained with the standard method, but the error bars are
such that the count rates are consistent with each other. Clearly, to
obtain the correct {\it ASCA}/SIS count rates for SAX J1808.4--3658,
it is very important to understand the local background level.

Using the MLR method, we were able to extract a spectrum of the
source. However, the spectrum is of low statistical quality and
multiple models could be fitted to the data equally well. The spectrum
can be fitted using a blackbody model (resulting in $kT$ of $\sim$0.65
keV) or with a power-law model (with a photon index of $\sim1.7$).  We
have also investigated the effects of the uncertainties in the column
density\footnote{We have reanalyzed the brightest burst seen with the
{\it BeppoSAX} Wide Field Camera's during the 1998 outburst of SAX
J1808.4--3658 (in 't Zand et al. 1998, 2001) and we determined that
the column density should be $< 5\times10^{21}$ cm$^{-2}$.}  towards
SAX J1808.4--3658 by fitting the spectrum with an $N_{\rm H}$ between
0.6 and 6 $\times 10^{21}$ cm$^{-2}$.  A range of fluxes was obtained
and we quote the typical unabsorbed fluxes, which include statistical
errors (as due to the fit procedure) and systematic errors (i.e., the
uncertainties in the background and in the column density; using
different spectral shapes like black-body or power-law): 7\pp4 (0.5--5
keV) and 10\pp6 $\times 10^{-14}$ \funit~(0.5--10 keV). Within the
errors, those fluxes are consistent with what has been reported
previously by Dotani et al. (2000).

At the position of the {\it BeppoSAX} source, no source could be
detected, with a 3$\sigma$ upper limit on the count rate of
$1.5\times10^{-3}$ counts s$^{-1}$ (0.5--5.0 keV; single SIS;
depending on the spectral shape and the column density this would
result in an unabsorbed 0.5--10 keV flux range of $0.3-1 \times
10^{-13}$ \funit).  Two of the remaining three {\it BeppoSAX} sources
were detected during the {\it ASCA} observation
(Fig.~\ref{fig:ascasis_image}).  SAX J1809.0--3659 was both detected
in the SIS (although at the edge of the CCD detectors; 5$\sigma$ using
the MLR method) and GIS detectors (4.5$\sigma$). SAX J1808.2--3654 was
outside the field of view (FOV) of the SIS but inside that of the GIS
and was detected by this instrument (3.5$\sigma$). The positions of
those two sources are consistent with those obtained using {\it
BeppoSAX} (Tab.~\ref{tab:fov}).  SAX J1808.5--3703 was also outside
the FOV of the SIS. The GIS did not detect the source.  An additional
source was detected with the GIS at a position of R.A. = 18$^{\rm h}$
08$^{\rm m}$ 02.8$^{\rm s}$ and Dec. = --37\degr~02$'$ 41.7$''$
($>9\sigma$; designated AX J1808.0--3702), although it might not be a
point source but an extended source (note that this causes additional
systematical errors on the source position which can be as large as
2$'$). No pronounced counterpart of this source is visible in the {\it
BeppoSAX}/MECS image (Fig.~\ref{fig:image}), but it might be present
at a 3$\sigma$ level.

\section{Discussion}

We have obtained {\it BeppoSAX}/NFI observations of the millisecond
X-ray pulsar during its 2000 outburst. Due to the extreme variability
during this outburst (see Wijnands et al. 2001), the {\it BeppoSAX}
observations were unfortunately taken when the source had very low
luminosities.  We detected four field sources during our
observations. One source is 1.63$'$ away from the position of the
millisecond X-ray pulsar, considerably larger than the 1$\sigma$ error
of 0.55$'$. This source can only be identified with SAX J1808.4--3658
if we assume that the {\it BeppoSAX} positional reconstruction is not
completely understood. If indeed this source cannot be identified with
the millisecond X-ray pulsar, then it should be designated SAX
J1808.6--3658.

A possible non-detection of SAX J1808.4--3658 during our {\it
BeppoSAX} observations would result in an 3$\sigma$ upper limit on the
count rate of $1.1 \times 10^{-3}$ counts s$^{-1}$ (1.3--10 keV; both
MECS units active). When assuming an $N_{\rm H}$ of 1.22 $\times
10^{21}$ cm$^{-2}$ and a power-law shaped spectrum with index 2, this
would result in an upper limit on the luminosity of
$\sim1\times10^{32}$ \Lunit~(0.5--10 keV; for a distance of 2.5
kpc). In our {\it ASCA} observations we could not detect a source on
the best position of the {\it BeppoSAX} source with a range of upper
limits on its X-ray flux of $0.3-1.0\times10^{-13}$ \funit~(0.5--10.0
keV; unabsorbed). If the column density is indeed
$\sim1.22\times10^{21}$ cm$^{-1}$ then the upper limit determined with
{\it ASCA} ($<0.6 \times 10^{-13}$ \funit) was lower then flux
detected during the {\it BeppoSAX} observations (1.9\pp0.3
$\times10^{-13}$ \funit).  This discrepancy can mean that indeed the
{\it BeppoSAX} source is the millisecond pulsar or also this source is
variable. In the latter case, the fact that this source is detected
during all {\it BeppoSAX}/NFI observations (including the 1999 March
17) and not during the only {\it ASCA} observation should then be
regarded as a coincidence.  The {\it ASCA} fluxes obtained (using
W3PIMMS) for the other three {\it BeppoSAX} sources were consistent
with the values obtained from the {\it BeppoSAX} fluxes (taking into
account the uncertainties in the spectra, the internal absorption of
the sources, and in calculating the fluxes using W3PIMMS).  The nature
of those field sources is unknown, but they could be, e.g., AGN.

If indeed the {\it BeppoSAX} source is a previously unknown source,
its proximity to the millisecond pulsar raises the question if all the
outbursts so far observed from the pulsar are indeed due to this
source and not from this extra source.  It is clear that the source
which exhibited the 1998 and 2000 outburst was the millisecond X-ray
pulsar because during both outbursts the 401 Hz pulsations were
detected (Wijnands \& van der Klis 1998; van der Klis et
al. 2000). Furthermore, during both outbursts the optical counterpart
increased considerable in luminosity (Giles et al. 1999; Wachter \&
Hoard 2000; Wachter et al. 2000; note, that this secures the
sub-arcsecond accurate position of the pulsar). However, the {\it
BeppoSAX}/WFC positional errors of the persistent source detected
during the 1996 outburst are consistent with both the position of the
millisecond pulsar and the {\it BeppoSAX} source.  Still, there are
two measurements which provide convincing evidence that the 1996
outburst was also due to the millisecond X-ray pulsar. First, the 99\%
error regions of the three bursts detected during that outburst are
all consistent with the pulsar, while that of one burst is clearly
inconsistent with the other source. We note that all three bursts were
observed in a small time window during the 1996 outburst and are most
likely due to same source. Second, there is a marginal detection of a
401 Hz oscillation during one of the three bursts (In 't Zand et
al. 2000).

Independent whether the {\it BeppoSAX} source can be identified with
the millisecond X-ray pulsar or that it is an unrelated field source,
it is clear that SAX J1808.4--3658 has very low luminosities in
quiescence.  As discussed in detail by Wijnands et al. (2001), the
very low luminosity of SAX J1808.4--3658 during the {\it BeppoSAX}
observations demonstrate the highly variable nature of the source
during its 2000 outburst. For example, the source luminosity was below
a few times $10^{32}$ \Lunit~on 2000 March 5--6 and 8, it reached
$\sim10^{35}$ \Lunit~on 2000 March 11 as observed with the {\it
RXTE}/PCA, but the source was very dim again with {\it BeppoSAX}/NFI
on 2000 March 22--23. Also, on several occasions after 2000 March the
source could again be detected with the {\it RXTE}/PCA. Therefore,
although the X-ray luminosity was very low during our {\it BeppoSAX}
observations, the clear long-term activity of the source demonstrated
that these observations cannot be regarded as true quiescent
observations.

We reanalyzed the {\it ASCA} observation of SAX J1808.4--3658 reported
by Dotani et al. (2000). We confirm the detection of the millisecond
X-ray pulsar. Due to the uncertainties in its exact {\it ASCA} count
rate (see section~\ref{section:asca}) and its exact spectral shape,
the exact quiescent luminosity of the millisecond pulsar is difficult
to constrain but it is most likely in the range 0.3--1$\times10^{32}$
\Lunit~(0.5--10 keV). Despite this large uncertainty in its quiescent
luminosity, the source is clearly fainter than the other neutron-star
LMXB X-ray transients when they are in quiescence (which are larger
than a few times $10^{32-33}$ \Lunit). If the quiescent luminosity of
SAX J1808.4--3658 is indeed as low as a few times $10^{31}$ \Lunit,
then it is unclear whether the found distinction (e.g., Garcia et
al. 2001 and references therein) between the quiescent luminosity of
neutron-star transients and black-hole transients (a few times
$10^{30-31}$ \Lunit) can hold. If certain neutron-star X-ray
transients can indeed be as dim in quiescence as the black-hole
systems, then the reasons to invoke the presence of event horizons in
the black-hole systems will disappear (Garcia et al. 2001). More
observations of neutron-star transients in quiescence are required to
assess the validity of reported quiescent luminosity difference
between the two types of systems. Such observations will also give
important information to assess the question whether or not the
millisecond X-ray pulsar is not only unique with respect to its
pulsating nature, but also with respect to its quiescent behavior.

Brown, Bildsten, \& Rutledge (1998) proposed that the quiescent X-ray
luminosities of neutron-star LMXB transients depend on their
time-averaged mass accretion rates. Compared with the other systems,
the time-averaged accretion rate of the millisecond X-ray pulsar is
low and Brown et al. (1998) suggested that this source should have a
low quiescent X-ray luminosity. Within the uncertainties of their
model and our observations, the predicted quiescent luminosity is
consistent with the detected one. It is also possible that this low
quiescent luminosity is related to the fact that the source is so far
the only accretion-driven millisecond X-ray pulsar known. For example,
if this uniqueness is due to a higher neutron-star magnetic field than
that of the other sources, then during quiescence the so-called
propeller mechanism might inhibit residual accretion onto the neutron
star. Indications for residual accretion in quiescence have been found
for the other neutron-star systems, which is mainly inferred from the
power-law tail in their quiescent X-ray spectrum and the variability
of several sources in quiescence (however, see Ushomirsky \& Rutledge
2001 for a discussion about X-ray variability due to deep crustal
heating). However, only at most $\sim$50\% of the quiescent X-ray flux
of those sources is due to this power-law component, suggesting that
the low quiescent luminosity of SAX J1808.4--3658 can only in part be
due to the lack of residual accretion in this system.

It is very well possible that the very low quiescent luminosity is due
to a combination of the two above mentioned effects. The low
time-averaged accretion rate of SAX J1808.4--3658 will not heat-up its
neutron star to the same temperature as in the other systems, causing
a lower thermal luminosity from its surface. Also, when residual
accretion is inhibited due to its magnetic field, the power-law
component is absent and the total quiescent X-ray luminosity of SAX
J1808.4--3658 could be very low compared to the other neutron-star
X-ray transients in quiescence.  High quality quiescent X-ray spectra
(e.g., with {\it XMM-Newton} or {\it Chandra}) of SAX J1808.4--3658
will help considerably to understand its low quiescent luminosity.

{\it Note added in manuscript:} After submission of our paper, Campana
et al. (2001) reported preliminary results on their {\it XMM-Newton}
observation of SAX J1808.4--3658 in quiescence. They detected the
source at a luminosity of approximately $5\times10^{31}$
\Lunit~(0.5--10 keV), consistent with our quiescent {\it ASCA}
luminosity. They also reported that at the position of the {\it
BeppoSAX} source no source could be detected. However, the {\it
XMM-Newton} observation was not taken at the same time as our {\it
BeppoSAX} observations during the 2000 outburst and variability of
this {\it BeppoSAX} source cannot be ruled out. Analysis of the DDT
{\it XMM-Newton} observation taken on 8 March 2000, will definitely
settled this issue because it was taken very close in time to our {\it
BeppoSAX} observations.

\acknowledgments

This work was supported by NASA through Chandra Postdoctoral
Fellowship grant number PF9-10010 awarded by CXC, which is operated by
SAO for NASA under contract NAS8-39073. MK acknowledges support by the
Netherlands Organization for Scientific Research (NWO).  We thank Jon
Miller and Erik Kuulkers for helpful discussions and carefully reading
a previous version of this paper. We thank the referee, Josh Grindlay,
for useful comments on our paper.

\hoffset -1cm

\begin{deluxetable}{ccc}
\tablecolumns{3}
\tablewidth{0pt}
\tablecaption{Log of the observations \label{tab:log}}
\tablehead{
Satellite      & Day of observation    & Source state}
\startdata
{\it BeppoSAX} & 17--19 March 1999     & Quiescence\\
{\it ASCA}     & 17--20 September 1999 & Quiescence\\
{\it BeppoSAX} & 5--6  March 2000      & 2000 outburst\\
               & 8 March 2000          & \\
               & 22--22 March 2000     &\\
\enddata
\end{deluxetable}

\begin{deluxetable}{cccccc}
\tabletypesize{\tiny}
\tablecolumns{6}
\tablewidth{0pt}
\tablecaption{Sources detected in the {\it BeppoSAX} and {\it ASCA} observations \label{tab:fov}}
\tablehead{
Source                & Satellite          & \multicolumn{3}{c}{Position$^a$}                                                & Count rate$^b$  \\
                      &                    & R.A.                                      & Dec.                     & Error    & ($10^{-3}$ counts s$^{-1}$)} 
\startdata														\\
SAX J1808.4--3658     & Optical            & 18$^{\rm h}$ 08$^{\rm m}$ 27.54$^{\rm s}$ & --36\degr~58$'$ 44.3$''$ & 0.2$''$  & \\
                      & {\it ASCA}/SIS     & 18$^{\rm h}$ 08$^{\rm m}$ 26.9$^{\rm s}$  & --36\degr~58$'$ 35.9$''$ & 0.24$'$  & 2.4\pp0.3\\
                      &                    &                                           &                          &          & 1.6\pp0.3\\
                      & {\it ASCA}/GIS     & 18$^{\rm h}$ 08$^{\rm m}$ 29.8$^{\rm s}$  & --36\degr~58$'$ 27.6$''$ & 0.57$'$  & \\
{\it BeppoSAX} source & {\it BeppoSAX}     & 18$^{\rm h}$ 08$^{\rm m}$ 35.1$^{\rm s}$  & --36\degr~58$'$ 07.8$''$ & 0.55$'$  & 1.9\pp0.3\\
SAX J1809.0--3659     & {\it BeppoSAX}     & 18$^{\rm h}$ 09$^{\rm m}$ 02.9$^{\rm s}$  & --36\degr~59$'$ 42.3$''$ & 0.56$'$  & 1.9\pp0.3\\
                      & {\it ASCA}/SIS     & 18$^{\rm h}$ 09$^{\rm m}$ 02.3$^{\rm s}$  & --36\degr~59$'$ 39.3$''$ & 0.23$'$  & 2.8\pp0.5\\
                      & {\it ASCA}/GIS     & 18$^{\rm h}$ 09$^{\rm m}$ 02.1$^{\rm s}$  & --36\degr~59$'$ 19.5$''$ & 0.53$'$  & \\ 
SAX J1808.2--3654     & {\it BeppoSAX}     & 18$^{\rm h}$ 08$^{\rm m}$ 12.9$^{\rm s}$  & --36\degr~54$'$ 42.7$''$ & 0.59$'$  & 1.6\pp0.2\\
                      & {\it ASCA}/GIS     & 18$^{\rm h}$ 08$^{\rm m}$ 10.3$^{\rm s}$  & --36\degr~54$'$ 11.3$''$ & 0.63$'$  & 1.2\pp0.5\\
SAX J1808.5--3703     & {\it BeppoSAX}     & 18$^{\rm h}$ 08$^{\rm m}$ 32.0$^{\rm s}$  & --37\degr~03$'$ 03.1$''$ & 0.73$'$  & 0.8\pp0.2\\
                      & {\it ASCA}/GIS     &                                           &                          &          & $<$1.6\\
\enddata 

\tablenotetext{a}{Source positions (J2000.0) as determined with {\it
BeppoSAX} or {\it ASCA} and for SAX J1808.4--3658 also from optical
observations (Giles et al. 1999). The error radii on the positions are
for 68\% confidence levels and are composed of the direct sum of the
statistical plus systematical errors.  The {\it ASCA} positions were
determined using the combined SIS-0 and SIS-1 data for SAX
J1808.4--3658 and SAX J1809.0--3659, and the combined GIS-2 and GIS-3
data for SAX J1808.2--3654.}

\tablenotetext{b}{Count rates for the photon energy range 1.3--10 keV
({\it BeppoSAX}, MECS, both active units), 0.5--5.0 keV ({\it ASCA},
single SIS), or 0.7--5.0 keV ({\it ASCA}, single GIS). The errors on
the count rates are for 1$\sigma$ and the upper limit for
3$\sigma$. For SAX J1808.4--3658 the first SIS row is for assuming a
flat background distribution and for the second one the background is
modeled using the three extra point sources (see
section~\ref{section:asca}). When SIS count rates are available no GIS
ones are give because the SIS ones are more reliable.}

\end{deluxetable}

\newpage

\begin{references}

\reference{}Asai, K., Dotani, T., Mitsuda, K., Hoshi, R., Vaughan, B.,
Tanaka, Y., Inoue, H. 1996, \pasj, 48, 257

\reference{}Asai, K., Dotani, T., Hoshi, R., Tanaka, Y., Robison,
C. R., Terada, K. 1998, \pasj, 50, 611

\reference{}Brown, E. F., Bildsten, L., \& Rutledge, R. E. 1998, \apj,
504, l95

\reference{}Campana, S., Stella, L., Gastaldell, F., Israel, G.,
Mereghetti, S., Burderi, L., Di Salvo, T. 2001, Poster presented at
``New Visions of the X-ray Universe in the XMM-Newton and Chandra
era'', 26--30 November 2001, Noordwijk, The Netherlands


\reference{}Dotani, T., Asai, K., \& Wijnands, R. 2000, \apj, 543,
L145

\reference{}Gaensler, B. M., Stappers, B. W., \& Getts, T. J. 1999,
\apj, 522, L117

\reference{}Garcia, M. R., McClintock, J. E., Narayan, R., Callanan,
P., Murray, S. S. 2001, \apj, 553, L47

\reference{}Gilfanov, M., Revnivtsev, M., Sunyaev, R., Churazov,
E. 1998, \aap, 338, L83

\reference{}Giles, A. B., Hill, K. M., Greenhill, J. G. 1999, \mnras,
304, 47

\reference{}Gotthelf, E. V., Ueda, Y., Fujimoto, R., Kii, T., Yamaoka,
K. 2000, \apj, 543, 417

\reference{}In 't Zand, J. J. M., Heise, J., Muller, J. M., Bazzano,
A., Cocchi, M., Nataluccci, L., Ubertinit, P. 1998, \aap, 331, L251

\reference{}In 't Zand, J. J. M., Kuiper, L., Amati, L., Antonelli,
L. A., Hurley, K. et al. 2000, \apj, 545, 266

\reference{}In 't Zand, J. J. M., Cornelisse, R., Kuulkers, E.,
Kuiper, L., Bazzano, A., Cocchi, M., Heise, J., Muller, J. M.,
Natalucci, L., Smith, M. J. S., Ubertinit, P.  2001 \aap, 372, 916


\reference{}Kuiper, L., Hermsen, W., Bennett, K., Carrami\~nana,
L. A., McConnell, M., Sch\"onfelder, V. 1998, \aap, 337, 421
       
\reference{}Kuulkers, E., et al. 2000, \apj, 538, 638

\reference{}Marshall, F. E. 1998, \iaucirc 6876
       
\reference{}Predehl, P. \& Schmitt, J. H. M. M. 1995, \aap, 293, 889

\reference{}Stella, L., Campana, S., Mereghetti, S., Ricci, D.,
Israel, G. L. 2000, \apj, 537, L115
       
\reference{}Ushomirsky, G. \& Rutledge, R. E 2001, \mnras, 325, 1157

\reference{}van der Klis, M., Chakrabarty, D., Lee, J. C., Morgan,
E. H., Wijnands, R., Markwardt, C. B., Swank, J. H. 2000, \iaucirc 7358

\reference{}Wachter, S. \& Hoard, D. W. 2000, \iaucirc, 7363

\reference{}Wachter, S., Hoard, D. W., Bailyn, C., Jain, R., Kaaret,
P., Corbel, S., Wijnands, R. 2000, HEAD 32, 24.15

\reference{}Wang, Z. et al. 2001, \apj, 563, L61

\reference{}Wijnands, R. \& van der Klis, M. 1998, Nature, 394, 344

\reference{}Wijnands, R., Chakrabarty, D., Morgan, E., van der Klis,
M. 2000, \iaucirc 7369

\reference{}Wijnands, R., M\'endez, M., Markwardt, C., van der Klis,
M., Chakrabarty, D., Morgan, E., 2001, \apj, 560, 892

\end{references}
\end{document}